\documentclass[aps,prd, notitlepage, onecolumn,superscriptaddress,floatfix,letterpaper,nofootinbib, longbibliography]{revtex4-1}

\usepackage{amsmath, amsthm, amssymb,slashed}

\usepackage[usenames, dvipsnames]{color}
\usepackage[svgnames]{xcolor}
\usepackage[colorlinks,citecolor=RoyalBlue, urlcolor=RoyalBlue, linkcolor=RoyalBlue ]{hyperref} 

\usepackage[normalem]{ulem}

\usepackage{booktabs}

\definecolor{mygray}{gray}{0.6}

\usepackage{upgreek}
\usepackage{bbm}

\newenvironment{myfont}[2][]{\csname#2\endcsname[#1]}{}

\usepackage{slashed}
\usepackage[makeroom]{cancel}
\usepackage[normalem]{ulem}
\usepackage{soul}
\newcommand{\stkout}[1]{\ifmmode\text{\sout{\ensuremath{#1}}}\else\sout{#1}\fi}

\usepackage{sseq}
\usepackage[all,cmtip]{xy}
\usepackage{tikz-cd}
\usepackage{tikz}
\usetikzlibrary{matrix}
\usetikzlibrary{decorations.markings}
\usetikzlibrary{tikzmark,decorations.pathreplacing,positioning}
\usepackage{amsfonts}
\usepackage{multirow}

\newcommand{\bea}{\begin{eqnarray}}
\newcommand{\eea}{\end{eqnarray}}
\def\be{\begin{equation}}
\def\ee{\end{equation}}

\newcommand{\ii}{\hspace{1pt}\mathrm{i}\hspace{1pt}}

\definecolor{red}{rgb}{1,0,0}
\definecolor{blue}{rgb}{0,0,1}
\definecolor{dblue}{rgb}{0,0,0.4}
\definecolor{green}{rgb}{0,1,0}
\definecolor{black}{rgb}{0,0,0}
\definecolor{white}{rgb}{1,1,1}

\definecolor{brn}{rgb}{.8,.4,.0}
\definecolor{redo}{rgb}{1,.5,.0}
\definecolor{ddgrn}{rgb}{0,0.4,0}
\definecolor{dgrn}{rgb}{0,0.55,0}
\definecolor{dbl}{rgb}{0,0,0.5}

\usepackage[bbgreekl]{mathbbol}
\usepackage{amscd}

\newcommand{\Z}{\mathbb{Z}}

\newcommand{\R}{\mathbb{R}}

\newcommand{\dd}{\mathrm{d}}

\newcommand{\Refe}[1]{Ref.~[\onlinecite{#1}]}
 
\newcommand{\eq}[1]{eq.~(\ref{#1})}

\newcommand{\ch}{{\rm ch}}

\newcommand{\bpm}{\begin{pmatrix}}
\newcommand{\epm}{\end{pmatrix}}
\newcommand{\bmm}{\begin{matrix}}
\newcommand{\emm}{\end{matrix}}

\def\Z{{\mathbb{Z}}}

\def\R{{\mathbb{R}}}

\def \Hom{\operatorname{Hom}}

\def \H{\operatorname{H}}

\def \Z{\mathbb{Z}}

\newcommand {\emptycomment}[1]{}

\def\B{\mathrm{B}}

\newcommand{\Spin}{{\rm Spin}}
\newcommand{\U}{{\rm U}}
\newcommand{\SU}{{\rm SU}}

\newcommand{\Pin}{{\rm Pin}}

\usepackage{centernot}
\newcommand{\nn}{{\nonumber}}
\newcommand{\Sec}[1]{Sec.~\ref{#1}} 

\usepackage{enumitem} 
\usepackage{mathtools,amssymb,varwidth}

\newcommand{\Fig}[1]{Fig.~\ref{#1}} 
\newcommand{\Table}[1]{Table \ref{#1}}

\usepackage{datetime}

\newcommand{\rF}{{\rm F}}

\newcommand{\GL}{{\rm GL}}

\newcommand{\rA}{{\rm A}}

\def\ch{\mathrm{ch}}

\usepackage{array}

\begin{document}


\title{Topological 
Quantum Dark Matter via Global
Anomaly Cancellation
}

\author{Juven Wang}
\affiliation{London Institute for Mathematical Sciences, Royal Institution, W1S 4BS, UK}
\affiliation{Center of Mathematical Sciences and Applications, Harvard University, MA 02138, USA}


\begin{abstract}

Standard Model (SM) with 15 Weyl fermions per family (lacking the 16th, the sterile right-handed neutrino $\nu_R$) suffers from mixed gauge-gravitational anomalies tied to baryon number plus or minus lepton number ${\bf B} \pm {\bf L}$ symmetry. Including $\nu_R$ per family can cancel these anomalies, but when ${\bf B} \pm {\bf L}$ symmetry is preserved as discrete finite subgroups rather than a continuous U(1), the perturbative local anomalies become nonperturbative global anomalies. In this work, we systematically enumerate these gauge-gravitational global anomalies involving discrete ${\bf B} \pm {\bf L}$ that are enhanced from the fermion parity $\mathbb{Z}_2^{\rm F}$ to $\mathbb{Z}_{2N}^{\rm F}$, with $N=2,3,4,6,9$, etc. The discreteness of ${\bf B} \pm {\bf L}$ is constrained by multi-fermion deformations beyond-the-SM and the family number $N_f$. Unlike the free quadratic $\nu_R$ Majorana mass gap preserving the minimal $\mathbb{Z}_2^{\rm F}$, we explore novel scenarios canceling $({\bf B} \pm {\bf L})$-gravitational anomalies while preserving the $\mathbb{Z}_{2N}^{\rm F}$ discrete symmetries, featuring 4-dimensional interacting gapped topological orders (potentially with or without low-energy topological quantum field theory 
descriptions) or gapless sectors (e.g., conformal field theories). 
We propose symmetric anomalous sectors as quantum dark matter to cancel SM's global anomalies. 
We find the uniqueness of 
the family number at $N_f=3$,
such that when the representation of 
$\mathbb{Z}_{2N}^{\rm F}$
from the faithful ${\bf B} + {\bf L}$
for baryons
at $N=N_c=3$
is extended to the faithful ${\bf Q} + N_c {\bf L}$ for quarks at $N=N_c N_f=9$,
this symmetry extension 
$\mathbb{Z}_{N_c=3} \to \mathbb{Z}_{N_c N_f =9} \to \mathbb{Z}_{N_f =3} $
matches with the
topological order dark matter construction.
Key implications include: (1) a 5th force mediating between SM and dark matter via discrete ${\bf B} \pm {\bf L}$ gauge fields,  (2) dark matter as topological order quantum matter with gapped anyon excitations at ends of extended defects, and (3) Ultra Unification and topological leptogenesis.

\end{abstract}


\maketitle

\tableofcontents



\newpage
\section{Introduction and Summary}

Standard Model (SM), with 15 Weyl fermions per family and without 
the 16th Weyl fermion sterile right-handed neutrino $\nu_R$,
suffers from the perturbative local
mixed-gauge-gravitational anomalies 
\cite{Eguchi:1976db, AlvarezGaume1983igWitten1984,
Alvarez-Gaume:1984zlq}
between the lepton number ${\bf L}$ symmetry and gravitational background fields, in 3+1d or simply 4d spacetime. 
Namely these anomalies are computable
via perturbative triangle Feynman diagrams
$\U(1)_{\bf L}^3$ and $\U(1)_{\bf L}$-gravity$^2$, with the anomaly index coefficient $-N_f + n_{\nu_R}$,
counting the difference between the family or generation number $N_f$ (typically $N_f=3$) and the total right-hand neutrino number $n_{\nu_R}$. See recent related expositions about this anomaly index $-N_f + n_{\nu_R}$ for examples
in  \cite{JW2006.16996, JW2008.06499, JW2012.15860, WangWanYou2112.14765, WangWanYou2204.08393, Putrov:2023jqi2302.14862, Wang:2024auy2501.00607}.
However, because of
the analogous Adler-Bell-Jackiw anomalies
\cite{Adler1969gkABJ, Bell1969tsABJ}
via the SM electroweak gauge instanton
\cite{BelavinBPST1975, tHooft1976ripPRL, JackiwRebbi:1976pf, 
CallanDashenGross:1976je},
instead of thinking of the classical lepton number ${\bf L}$ symmetry,
only the baryon number plus or minus lepton number ${\bf B} \pm {\bf L}$ symmetries are physically meaningful quantum mechanical symmetries pertinent in the SM:\footnote{In fact, 
the $\U(1)_{{\bf B} - {\bf L}}$ is a faithful symmetry for the SM such that there exist gauge-invariant local operators that have a unit U(1) charge.
The $\U(1)_{{\bf Q} - N_c {\bf L}}$
is not a faithful symmetry for the SM 
because there exists no gauge-invariant local operator that has a unit U(1) charge;
but the gauge-invariant local operator has a minimal $N_c=3$ charge. But $\U(1)_{{\bf Q} - N_c {\bf L}}$ is a faithful symmetry for free quarks, because there exists a free quark of that unit U(1) charge.\\
Similarly, the $\Z_{2 N_f, {\bf B} + {\bf L}}$ is a faithful symmetry for the SM such that there exist gauge-invariant local operators that have a unit U(1) charge.
The $\Z_{2 N_c N_f, {\bf Q} + N_c {\bf L}}$
is not a faithful symmetry for the SM 
because there exists no gauge-invariant local operator that has a unit $\Z_{2 N_c N_f, {\bf Q} + N_c {\bf L}}$ charge;
but the gauge-invariant local operator has a minimal $N_c=3$ charge. But $\Z_{2 N_c N_f, {\bf Q} + N_c {\bf L}}$ is a faithful symmetry for free quarks, because there exists a free quark of that unit $\Z_{2 N_c N_f, {\bf Q} + N_c {\bf L}}$ charge.\\
We thank Yunqin Zheng for enlightening discussions on this issue.
}

\begin{enumerate}
\item 
For the ${\bf B} - {\bf L}$ symmetry, 
there is a full 
faithful $\U(1)_{{\bf B} - {\bf L}}$
symmetry for the gauge-invariant baryons,
or an unfaithful $\U(1)_{{\bf Q} - N_c {\bf L}}$
symmetry for the gauge-invariant baryons
(but $\U(1)_{{\bf Q} - N_c {\bf L}}$ is faithful for the free quarks),
 survived under the SM electroweak gauge instanton, see Table \ref{table:SM}.

\item
For the ${\bf B} + {\bf L}$ symmetry, 
there is a full 
faithful $\Z_{2 N_f, {\bf B} + {\bf L}}$
symmetry for the gauge-invariant baryons
or an unfaithful $\Z_{2 N_c N_f, {\bf Q} + N_c {\bf L}}$
symmetry for the gauge-invariant baryons
(but $\Z_{2 N_c N_f, {\bf Q} + N_c {\bf L}}$ is faithful for the free quarks),
 survived under the SM electroweak gauge instanton,
see Table \ref{table:SM}.
\end{enumerate}

\begin{table}[!h]
\begin{tabular}{|c |  c  | c | c |  c |  c | c  | c|}
\hline
& 
$\bar{d}_R$ & $l_L$ & $q_L$ & $\bar{u}_R$
& $\bar{e}_R= e_L^+$ & $\bar{\nu}_R= {\nu}_L $ &$\phi_H$\\
\hline
${\SU(3)}$ & $\overline{\mathbf{3}}$ & $\mathbf{1}$ & ${{\mathbf{3}}}$ & $\overline{\mathbf{3}}$ & $\mathbf{1}$ & $\mathbf{1}$ & $\mathbf{1}$\\
${\SU(2)}$ & $\mathbf{1}$ & $\mathbf{2}$ & $\mathbf{2}$  & $\mathbf{1}$ & $\mathbf{1}$ & $\mathbf{1}$ & $\mathbf{2}$\\
$\U(1)_{Y}$ & 1/3 & $-1/2$ & 1/6 & $-2/3$ & 1 & 0 & ${1}/{2}$\\
$\U(1)_{\tilde Y }$ & 2 & $-3$ & 1 & $-4$ & 6 & 0 & $3$\\
$\U(1)_{\rm{EM}}$ & 1/3 & 0 or $-1$ & 2/3 or $-1/3$ & $-2/3$ & 1 & 0 & 0\\
$\U(1)_{{ \mathbf{B}-  \mathbf{L}}}
=\U(1) ^\rF$ & $-1/3$ & $-1$ & $1/3$ & $-1/3$ & 1 & 1 & 0\\
$\U(1)_{{{\bf Q}} - {N_c}{\bf L}}
=\U(1) ^\rF$ & $-1$ & $-3$ & 1 & $-1$ & 3 & 3 & 0\\
$\U(1)_{X} =\U(1) ^\rF$ 
& $-3$ & $-3$ & 1 & 1 & 1 & 5 & $-2$\\
$\Z_{5,X}$ & 2 & 2 & 1 & 1 & 1 & 0 & $-2$\\
$\Z_{4,X}= \Z_4^\rF$ & 1 & 1 & 1 & 1 & 1 & 1 & $-2$\\
$\Z_{8,X}= \Z_8^\rF$ & 5 & 5 & 1 & 1 & 1 & 5 & $-2$\\
{\parbox{3.8cm}{ 
\vspace{2pt}
$\Z_{ 2N_f=6, {{\bf B}} + {\bf L}}
=  \Z_6^\rF$ \\
for $N_f=3$;\\
or $\Z_2^\rF, \Z_4^\rF$\\
for $N_f=1,2$\\
(broken from 
$\U(1)_{{{\bf B}} + {\bf L}}$).
\vspace{2pt}}}
& $-1/3$ & $1$ & $1/3$ & $-1/3$ & $-1$ & $-1$ & 0\\
{\parbox{3.8cm}{ 
\vspace{2pt}
$\Z_{ 2N_cN_f=18, {{\bf Q}} + {N_c}{\bf L}}
= \Z_{18}^\rF$ \\
for $N_f=3$;\\
or $\Z_6^\rF, \Z_{12}^\rF$\\
for $N_f=1,2$\\
(broken from 
$\U(1)_{{{\bf Q}} + {N_c}{\bf L}}$).
\vspace{2pt}}}
& $-1$ & $3$ & 1 & $-1$ & $-3$ & $-3$ & 0\\
$\Z_{2}^\rF$ & 1 & 1 & 1 & 1 & 1 & 1 & 0\\
\hline
\end{tabular}
\caption{Follow the convention in \cite{Putrov:2023jqi2302.14862}, we show the representations of quarks and leptons in terms of  
Weyl fermions 
in various internal symmetry groups.
Each fermion is a spin-$\frac{1}{2}$ Weyl spinor 
${\bf 2}_L$ representation {of} the spacetime symmetry group Spin(1,3).
Each fermion is written as a left-handed particle $\psi_L$ or a right-handed anti-particle $\ii \sigma_2 \psi_R^*$.
}
\label{table:SM}
\end{table}

\begin{figure}[h]
    \centering
 \includegraphics[width=6.8in]{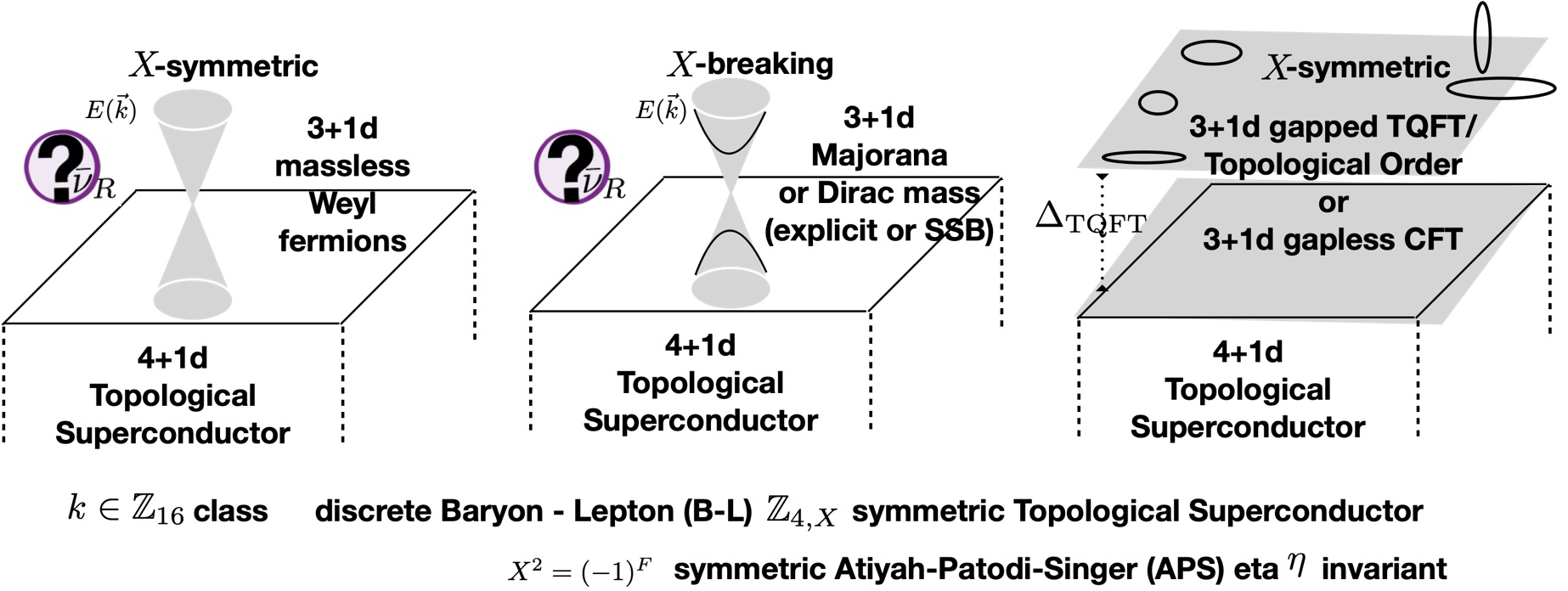}
    \caption{Here we give a specific example for $\Z_{16}$ class global anomaly cancellation for the mixed $\Z_{4,X}$-gravity anomalies between the SM and the BSM dark matter sector, 
    known for the model of Ultra Unification \cite{JW2006.16996, JW2008.06499, JW2012.15860, WangWanYou2112.14765, WangWanYou2204.08393}. Similar generalization for ${\bf B} \pm {\bf L}$ symmetry can be analogously obtained, too.\\
    (a) A single Weyl fermion with a unit charge $1 \in \Z_{4,X}$ can cancel an anomaly index $\upnu =1 \in \Z_{16}$.  
    A single Weyl fermion can live on the boundary of 5d $\Z_{16}$ class $\Z_{4,X}$-symmetric topological superconductor or SPTs in condensed matter, or known as 
    Atiyah-Patodi-Singer eta invariant
    or invertible topological field theory (iTFT) cobordism invariant \cite{GarciaEtxebarriaMontero2018ajm1808.00009, Hsieh2018ifc1808.02881, WW2019fxh1910.14668}. This can be the 16th Weyl fermion, the sterile right-handed neutrinos (denoted $\bar{\nu}_R$ so to be left-handed), shown in cartoon.\\
    (b) A single Weyl fermion is equivalent to a Majorana fermion in 4d,
    which can obtain a Majorana mass gap,
    but that Majorana mass term 
    breaks $\Z_{4,X}$ down to the 
    minimal fermion parity $\Z_2^{\rF}$.\\
    (c) The last scenario is inspired by the
    quantum condensed matter phenomena called the 2+1d \emph{symmetric anomalous boundary topological order} 
    on the boundary of 3+1d SPTs \cite{VishwanathSenthil1209.3058, Senthil1405.4015}.
    Here we consider its one higher-dimensional generalization:
    the 3+1d \emph{symmetric anomalous boundary topological order} 
    on the boundary of 4+1d SPTs.
    There are gapped extended line or surfaces defects in the 3+1d topological order.
    The open ends of extended defects that carry anyon statistics. 
    The last scenario could also include
    symmetric anomalous gapless sectors such as conformal field theories.
    }
    \label{fig:bulk-boundary-new-3+1d-4+1d-nuR}
\end{figure}

Although including the 16th Weyl fermion can cancel the SM's anomalies,
when the ${\bf B} \pm {\bf L}$ are preserved only as discrete finite subgroups instead of the conventional continuous U(1),
the perturbative local anomalies become 
nonperturbative global anomalies,\footnote{See the recent modern systematic studies of 
nonperturbative global anomalies in the context of the Standard Model in
Refs.~\cite{GarciaEtxebarriaMontero2018ajm1808.00009, Hsieh2018ifc1808.02881, DavighiGripaiosLohitsiri2019rcd1910.11277, WW2019fxh1910.14668}. For our terminology, we mean that:\\
$\bullet$ Perturbative local anomalies 
(e.g. \cite{AlvarezGaume1983igWitten1984,Alvarez-Gaume:1984zlq})
are detected by small (i.e. local)
continuous symmetry transformation connected to the identity transformation, hence it is sometimes called the continuous
anomaly.\\
$\bullet$ Non-perturbative global anomalies 
(e.g. \cite{Witten1985xe, Witten1982fp, WangWenWitten2018qoy1810.00844})
are detected by large (i.e.
global) discrete symmetry transformation disconnected from the
identity transformation, hence it is sometimes called the discrete
anomaly.
\\
$\bullet$ Sometimes the discrete anomaly is also used differently to describe the anomaly associated with discrete symmetries. 
See pioneer work in \cite{Ibanez1991hvRossPLB, Ibanez1992NPB, BanksDine1991xj9109045}.
In our case, we do have both the discrete global anomaly in both senses of (1) nonperturbative global anomalies, and (2) discrete ${\bf B} \pm {\bf L}$ symmetries.\\
$\bullet$ 't Hooft anomalies are referred to the anomaly of the global symmetry (that cannot be consistently dynamically gauged).\\
$\bullet$  Adler-Bell-Jackiw anomalies are the violations of the global symmetry in the presence of some (other) dynamically gauge fields. \\
$\bullet$ 
Dynamical gauge anomalies are meant to be canceled with dynamical gauge fields,
for the consistency under the name of anomaly cancellation.
\\
$\bullet$ Gauge or gravitational anomalies are anomalies associated with gauge or gravitational fields, that can be either background fields or dynamical fields.
Bosonic or fermionic anomalies
are anomalies associated with the anomalous boundary of one-higher dimensional bulk 
bosonic or fermionic Symmetry-Protected Topological states (SPTs).
}
there are alternative new scenarios that 
would cancel the $({\bf B} \pm {\bf L})$-gravitational anomaly 
but in such a novel way as to preserve the
discrete 
$({\bf B} \pm {\bf L})$-symmetry, see 
\Fig{fig:bulk-boundary-new-3+1d-4+1d-nuR}.

In this work, we systematically enumerate these 
gauge-gravitational global anomalies involving 
discrete ${\bf B} \pm {\bf L}$ that are enhanced 
from the fermion parity $\mathbb{Z}_2^{\rm F}$ to $\mathbb{Z}_{2N}^{\rm F}$,
with $N=2,3,4,6,9$, etc., see Appendix \ref{app:A}.
\footnote{The 
$\mathbb{Z}_2^{\rm F}$ is generated by $(-1)^\rF$ with the fermion number $\rF$,
so that $((-1)^\rF)^2=+1$.\\
The $\mathbb{Z}_{2N}^{\rm F} \supset \mathbb{Z}_2^{\rm F}$ contains the fermion parity as a normal subgroup such that
the quotient group $\mathbb{Z}_{2N}^{\rm F}/\mathbb{Z}_2^{\rm F}
=\mathbb{Z}_{N}$.\\
For $\mathbb{Z}_{2N,X}$ as $\mathbb{Z}_{2N}^{\rm F}$, we have
the generator $X^{2N}=+1$ as the cyclic group of order $2N$.
}
\begin{enumerate}
\item 
Phenomenologically, 
the discreteness of ${\bf B} - {\bf L}$ 
is \emph{either} imposed as a 
discretely dynamically gauged at the high-energy \cite{KraussWilczekPRLDiscrete1989}
\emph{or}
constrained by the 
allowed higher-order $2N$-body multi-fermion beyond-the-Standard-Model (BSM)  deformations. 
More precisely, we really require the 
$X$ symmetry  \cite{WeinbergPhysRevLett.43.1566, Wilczek1979hcZee, WilczekZeePLB1979}
\bea
 X \equiv 5({ \mathbf{B}-  \mathbf{L}})-\frac{2}{3} {\tilde Y}
  \equiv \frac{5}{N_C}({ \mathbf{Q}-  N_C \mathbf{L}})-\frac{2}{3} {\tilde Y}
\eea  
with the properly integer quantized hypercharge $\tilde Y$. See \Table{table:SM}.\\
\begin{itemize}
\item 
The $\Z_{4,X}$ has the advantage that 
all SM fermions have $\Z_{4,X}$ charge $1$,
thus we can 
consider the 4-body BSM multi-fermion $\Z_{4,X}$-preserving deformations  
\bea
\psi_q\psi_q\psi_q\psi_l, \quad \psi_{\bar q}\psi_q\psi_{\bar l}\psi_l, \quad 
\psi_{\bar q}\psi_q \psi_{\bar q}\psi_q, \quad \psi_{\bar l}\psi_l\psi_{\bar l}\psi_l.
\eea
because their $\Z_{4,X}$ charge is 
$4 = 0 \mod 4$.
\item 
The $\Z_{8,X}$ is less uniform so
SM quarks and leptons carry different 
$\Z_{8,X}$ charges. In addition to 4-body 
multi-fermion,
one can consider 8-body multi-fermion deformations
\bea
(\psi_{\bar q}\psi_q \psi_{\bar q}\psi_q)(\psi_{\bar q}\psi_q \psi_{\bar q}\psi_q),
\quad
(\psi_{\bar l}\psi_l\psi_{\bar l}\psi_l)
(\psi_{\bar l}\psi_l\psi_{\bar l}\psi_l),
\quad
(\psi_{\bar q}\psi_q \psi_{\bar q}\psi_q)
(\psi_{\bar l}\psi_l\psi_{\bar l}\psi_l), 
\quad
\dots
\eea
\end{itemize}
so that their $\Z_{8,X}$ charge is 
$0 \mod 8$.

Crucially, without any of 
those BSM deformations at the electroweak SM energy scale,
the ${\bf B} - {\bf L}$ or $X$ is preserved  as a full $\U(1)$ symmetry.

\item 
The discreteness of 
${\bf B} + {\bf L}$ 
is constrained by the family number $N_f$,
as faithful
$\Z_{2 N_f, {\bf B} + {\bf L}}$
or an unfaithful 
$\Z_{2 N_c N_f, {\bf Q} + N_c {\bf L}}$
symmetry for the SM,
for $N_c=3$. We will vary 
the family or generation number $N_f=1,2,3,4$, etc. to explore the uniqueness of $N_f=3$.

\end{enumerate}

\begin{figure}[!h]
    \centering
 \includegraphics[width=6.2in]{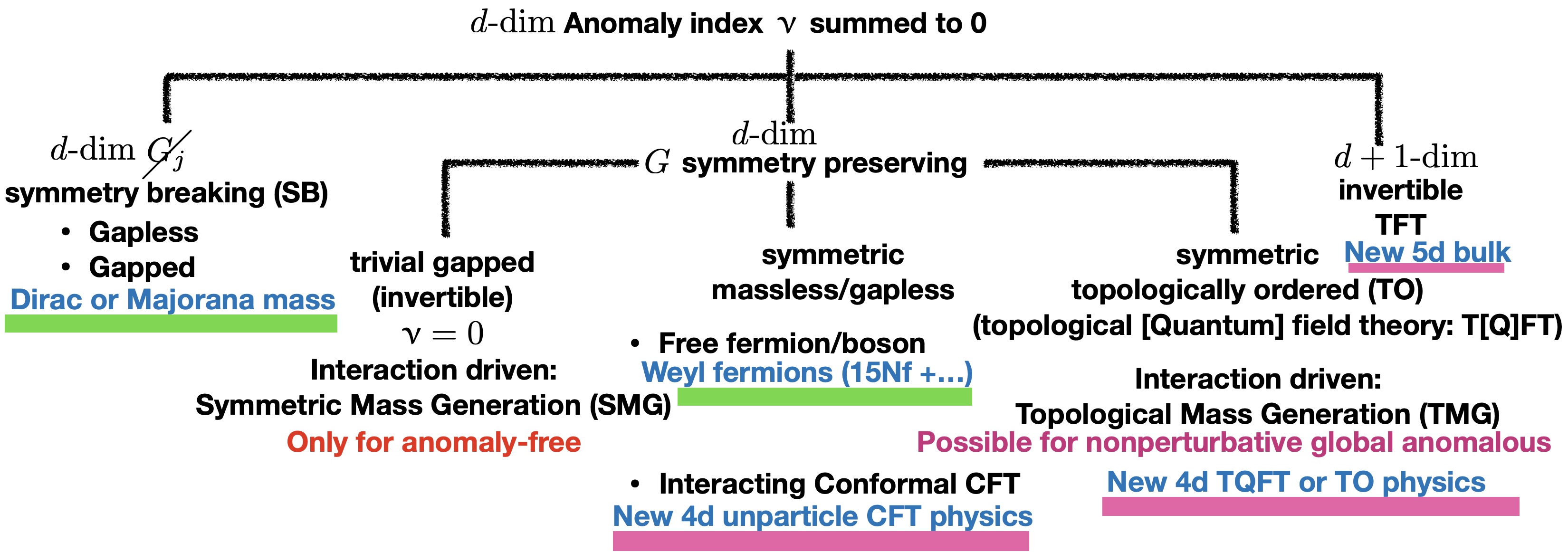}
    \caption{The anomaly cancellation demands the total summation of the anomaly index $\upnu = 0$.
    For perturbative local anomaly classified by an integer $\Z$ class, 
    the $\upnu = 0$ is strict.
    But for nonperturbative global anomaly
    classified by an integer $\Z_n$ class, 
    only the $\upnu = 0 \mod n$ is required.
    In addition to the familiar scenario of adding fermions (either gapless or gapped) to cancel the anomaly
    (marked in green), there are also novel scenarios (marked in pink) 
    include adding
    interacting 
    symmetric gapped topological order (TO)
    with or without low-energy 
    topological quantum field theory (TQFT) desciption,
    or symmetric gapless 
    conformal field theory (CFT),
    or an extra-dimensional bulk of invertible field theory.
    When the anomaly index is 
    $\upnu = 0 \mod n$, one can also use the
    Symmetric Mass Generation (SMG, e.g. see a review \cite{Wang:2022ucy2204.14271})
    mechanism to move between different quantum phases.} 
    \label{fig:UV-IR-anomaly-9-cancel-color}
\end{figure}

Unlike the free quadratic Majorana mass gap of $\nu_R$ preserving only 
the minimal fermion parity $\mathbb{Z}_2^{\rm F}$
(see \Fig{fig:bulk-boundary-new-3+1d-4+1d-nuR} (a) and (b)),
the exotic $\mathbb{Z}_{2N}^{\rm F}$-preserving
new scenarios can contain highly interacting 
symmetric anomalous gapped topological orders
(which inspiration originates from quantum condensed matter phenomena \cite{VishwanathSenthil1209.3058, Senthil1405.4015})
or symmetric anomalous gapless sectors such as conformal field theories, 
see \Fig{fig:bulk-boundary-new-3+1d-4+1d-nuR} (c).
These anomalous topological orders may or may not have topological quantum field theory (TQFT) descriptions at low energy
\cite{Yang:2023gvi2303.00719, Cheng:2024awi2411.05786}.
The enumerations of possible anomaly cancellation scenarios are summarized in \Fig{fig:UV-IR-anomaly-9-cancel-color}.

We propose introducing these symmetric anomalous sectors as 
quantum dark matter to cancel the SM's discrete gauge-gravitational global anomaly involving the discrete ${\bf B} \pm {\bf L}$.
We identify some of the mathematical constraints 
of the topologically ordered quantum dark matter.
The implications beyond the SM include:
\begin{enumerate}
\item The existence of the 5th force as the topological gauge force of
discrete ${\bf B} \pm {\bf L}$ gauge fields mediating between the SM 
and the dark matter. 
\item The dark matter can partly contain 
topological quantum matter,
such that the topologically ordered
gapped excitations at the open ends 
of extended line and surface defects can have anyon statistics, 
\item Ultra Unification \cite{JW2006.16996, JW2008.06499, JW2012.15860}
and topological leptogenesis \cite{Wang:2024auy2501.00607}.
\end{enumerate}

\begin{figure}[!h]
    \centering
 \includegraphics[width=5.8in]{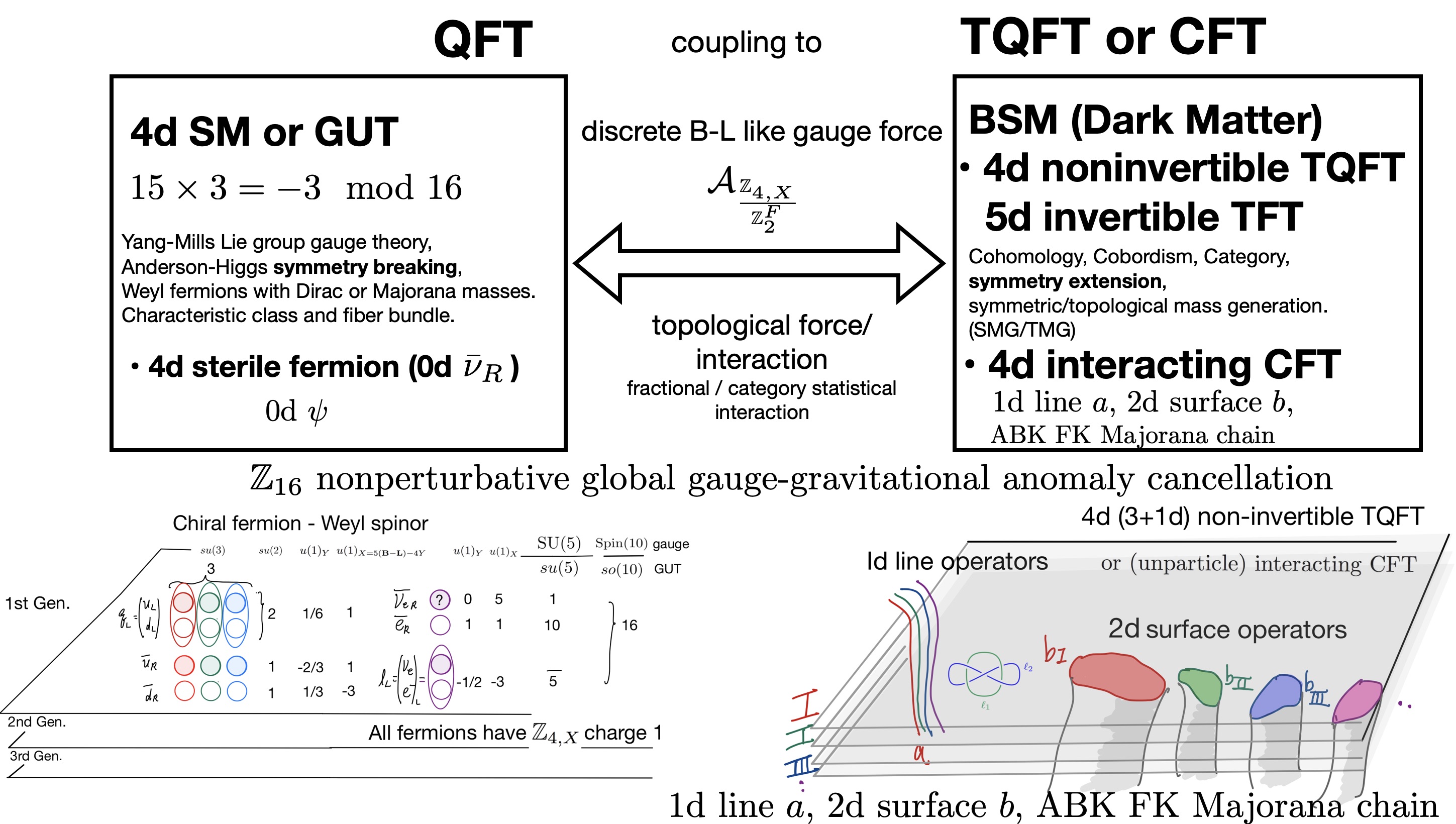}
    \caption{A schematic picture on the 
    nonperturbative global anomaly cancellation:
    how the
    quantum field theory (QFT) of the Standard Model (SM) can be coupled to the
    topological quantum field theory (TQFT)
    or conformal field theory (CFT) and others, via discrete gauge fields. 
    Shown here \Refe{JW2006.16996, JW2008.06499, JW2012.15860} considers discrete $X$ or
    ${\bf B} - {\bf L}$ gauge field.
    But more generally,
    for the SM,
    this works consider
    discrete
    ${\bf B} \pm {\bf L}$ gauge fields.
    This is a specific 
    QFT coupling to a TQFT scenario, see more explorations on this topic in \cite{Kapustin:2014gua1401.0740}.}
    \label{fig:QFT-TQFT-SM-BSM-full-new}
\end{figure}

\newpage

Below we will mainly use
the constraints obtained 
 in \cite{Cordova2019bsd1910.04962, Cordova1912.13069} or in
 3+1d boundary topological order 
of 4+1d SPTs obtained 
 in \cite{Yang:2023gvi2303.00719, Cheng:2024awi2411.05786} to comment about
the hypothetical
gapped or gapless topological quantum matter phases
as dark matter sectors.

\section{Discrete ${\bf B}+{\bf L}$ global anomaly and dark matter sector}

Based on the calculation done in Appendix \ref{app:A},
we find the following results.
(Since these anomalies listed below 
vanish for the
16-Weyl fermion SM, so we will only discuss the implications on the 15-Weyl fermion SM.)

For $\Z_{2 N_f, {\bf B} + {\bf L}}$
(faithful for free quarks), at $N_c=3$ color,

\begin{enumerate}
\item $N_f=1$, there is no global anomaly for 15-Weyl fermion SM.

\item $N_f=2$,
the anomaly index of the $N_f=2$-family SM 
(without 2 $\bar{\nu}_R$) is 
$(2 \mod 16).$
According to \cite{Cheng:2024awi2411.05786},
\begin{itemize}    
\item The $(2 \mod 16)$ corresponds to the 4d anomaly of beyond the group cohomology 5d SPT.
\item
For $\nu=N=2 \text{ or even } \in \Z_{16}$, we have a symmetric gapped $\Z_4$ gauge theory TQFT.
\item
For $\nu=N/2=1 \text{ or odd }\in \Z_{16}$,
we find a non-TQFT symmetric gapped state via stacking lower-dimensional (2+1)d non-discrete-gauge-theory topological order (that has TQFT descriptions) inhomogeneously.
\end{itemize}

\item $N_f=3$,
the anomaly index of the $N_f=3$-family SM 
(without 3 $\bar{\nu}_R$) is 
$(3 \mod 9).$
The $(3 \mod 9)$ corresponds to the 4d anomaly of the group cohomology 5d SPT,
see Appendix \ref{sec:Z9-fSPT}.

\end{enumerate}

For $\Z_{2 N_c N_f, {\bf Q} + N_c {\bf L}}$
(unfaithful for baryons, but faithful for free quarks or the ungauged SM without gauging the [SU(3)] color), at $N_c=3$ color,

\begin{enumerate}

\item $N_f=1$, the anomaly index is $0 \mod 9$ Thus, there is no needed to trivialize by $\Z_{N_c=3}$-extension via the construction in \cite{Wang2017locWWW1705.06728}.

\item $N_f=2$,
the anomaly index of the 2-family SM 
(without 2 $\bar{\nu}_R$) is 
$
(-2 \mod 16, \quad 0 \mod 9),
$
thus anomalous.
The $\Z_{N_c=3}$-extension via the construction in \cite{Wang2017locWWW1705.06728} cannot trivialize this anomaly.

\item $N_f=3$,
the anomaly index of the $N_f=3$-family SM 
(without 3 $\bar{\nu}_R$) is 
$(0 \mod 27, \quad 0 \mod 3).$

\begin{itemize}

\item This shows the uniqueness of $N_f=3$ or a multiple of 3 such that the following group extension 
\bea
1 \to \Z_{N_c=3} \to \Z_{N_cN_f=9}
\to \Z_{N_f=3} \to 1,
\eea
can trivialize the previous
$(3 \mod 9)$ anomaly for 
$\Z_{2 N_f=6, {\bf B} + {\bf L}}$.
Thus this also means that such a symmetry-extension can help to construct the symmetric gapped boundary topological order via the construction in \cite{Wang2017locWWW1705.06728}.
See more discussions on the uniqueness of
the $\Z_3$ symmetry
in Appendix \ref{sec:Z9-fSPT}.
\item 
Note that this $N_f=3$ family argument is different from the dimensional reduction argument via $c_-=24$ modular invariant in 2d CFT or framing anomaly free in 3d TQFT, and pure gravitational anomaly vanishing argument given in
\cite{Wang:2023tbj2312.14928}.

\item Note that our $\Z_3$ symmetry
assignment here only relies on the 
${\bf B} + {\bf L}$ symmetry of the standard SM,
thus crucially very different from
 the minimal-supersymmetric Standard Model (MSSM) previously
discussed in \cite{GarciaEtxebarriaMontero2018ajm1808.00009, Dreiner2005rd0512163}.

\item There are other physical motivations to consider the $\Z_{2 N_c N_f={18}, {\bf Q} + N_c {\bf L}}$ as a global or 
gauge symmetry in
proton stability  
\cite{KorenProtonStability2204.01741, WangWanYou2204.08393}
or in
cosmological lithium problem \cite{Koren:2022axd2204.01750}.
 
\end{itemize}

\end{enumerate}

\section{Discrete ${\bf B}-{\bf L}$ global anomaly and dark matter sector}

We consider $\Z_{2N,X}$ symmetry for the  
discrete ${\bf B}-{\bf L}$ symmetry,
\begin{enumerate}
\item $\Z_{4,X}$: 
The anomaly index of the 3-family SM 
(without 3 $\bar{\nu}_R$) is 
$
(-3 \mod 16).
$
Using a $\nu=\text{even} \in \Z_{16}$ index 
symmetric gapped $\Z_4$ gauge theory TQFT \cite{Cheng:2024awi2411.05786} is not enough to cancel the SM's $N_f=3$ anomaly, either some $\nu_R$, or some 4d gapless vector, or some extra dimensional 5d bulk, or some non-TQFT kind of fracton topological order is needed.

\item $\Z_{8,X}$: 
Missing 3 right-handed neutrinos (not found in the SM), each has $\Z_{8,X}$ charge $q=5$, thus the 3 right-handed neutrinos has 
$
3(-7 \mod 32, \quad 1 \mod 2) =(11 \mod 32, \quad 1 \mod 2).
$
The complementary anomaly index of the SM is 
$
(21 \mod 32, \quad 1 \mod 2).
$
Using a $\nu=\text{4k} \in \Z_{32}$ index 
symmetric gapped $\Z_4$ gauge theory TQFT \cite{Cheng:2024awi2411.05786} is not enough to cancel the SM's $N_f=3$ anomaly, either some $\nu_R$, or some 4d gapless vector, or some extra dimensional 5d bulk, or some non-TQFT kind of fracton topological order is needed.

\end{enumerate}

\section*{Acknowledgements}

JW thanks Meng Cheng, Pavel Putrov, Zheyan Wan, Matthew Yu, 
and Yunqin Zheng for helpful discussions.
JW is supported by LIMS and Ben Delo Fellowshop.
JW thanks the collaborators in the project for related discussions \cite{Wan:2024kaf}.

\appendix

\section{Anomaly of 3+1d Weyl fermion with discrete charge $q \in \Z_n$}
\label{app:A}

In this Appendix, for a single Weyl fermion theory 
assigned with a discrete charge $q \in \Z_n$ in 3+1d (or simply 4d) spacetime,
we derive the anomaly index, and relate the index to
a 4+1d Symmetry-Protected Topological state (SPTs),
or a cobordism invariant
(that is an invertible  field theory [iFT] or
an invertible topological field theory [iTFT]). 
We will follow
the discussion in the Appendix of
\cite{Putrov:2023jqi2302.14862}.
We will also require information from 
\cite{Hsieh2018ifc1808.02881}. 
We will recombine the linear combination of the generators
of the 3+1d anomaly (or 4+1d cobordism invariant) in \cite{Hsieh2018ifc1808.02881}
to derive the appropriate generator that can match with that of 
a charge $q$ Weyl fermion.

Below we may simply denote the 3+1d spacetime as 4d,
and 4+1d spacetime as 5d.

\subsection{Anomaly Polynomial}

We read the 4d Weyl fermion anomaly and its associated 5d invertible theory
action
$S_5= 2 \pi \int_{ M^5} I_5 \in 2 \pi \R$
from the 6d anomaly polynomial 
$I_6=\dd I_5$ whose integration over a closed 6-manifold is valued in $\Z$,
from
the $\hat{\rA}$ genus and the Chern character $\ch(\mathcal{E})$,
\bea
\hat{\rA} \, \ch(\mathcal{E}),
\eea
where $\hat{\rA}$ and $\ch(\mathcal{E})$ are given as:
\bea
\hat{\rA} &=&1-\frac{p_1}{24}+ \frac{7 p_1^2 - 4 p_2}{5760}+ \ldots, \\
\ch(\mathcal{E})&=&\mathrm{rank}\,\mathcal{E}+c_1(\mathcal{E})+
    \frac{1}{2}\left(c_1^2(\mathcal{E})-2c_2(\mathcal{E})\right)+
    \frac{1}{6}\left(
c_1^3(\mathcal{E})-3c_1(\mathcal{E})c_2(\mathcal{E})+3c_3(\mathcal{E})
    \right)+\ldots
\eea
{For a single left-handed Weyl fermion of charge $q$, we take $\mathcal{E}$ to be the complex line bundle associated with the corresponding representation of $\U(1)$}.
Hence, the fermionic 6d anomaly polynomial is integer quantized over a closed 6-manifold
\bea \label{eq:I6f}
\int_{M^6} I_{6,f} =
\int_{M^6} [\hat{\rA} \, \ch(\mathcal{E})]_6=
\int_{M^6} q^3 \frac{c_1^3}{6} -  
q\frac{c_1p_1}{24} \in \Z,
\eea
computed as the index of the 6d Dirac operator via Atiyah-Singer index theorem \cite{AlvarezGaume1983igWitten1984,Alvarez-Gaume:1984zlq}.
The 5d 
invertible theory
action is
\bea
S_5= 2 \pi \int_{ M^5} I_5 
=\int_{M^6}  q^3 A \frac{c_1^2}{6} -  
 q A \frac{p_1}{24} 
\in 2 \pi \R.
\eea
Next, we will start with two compatible 
$\Spin \times \U(1)$ and $\Spin \times_{\Z_2^\rF} \U(1) \equiv \Spin^c$ structures
for the fermion case with a U(1) symmetry,
to match with the appropriate generators of the cobordism group
$\Hom(\Omega_6^{\Spin^c},\Z)=\Z^2$
and $\Hom(\Omega_6^{\Spin \times \U(1)},\Z)=\Z^2$.
Then we consider the Weyl fermion with a charge $q$ of a discrete subgroup $\Z_n$ symmetry out of this $\U(1)$.

\subsection{${\Spin \times \U(1)}$ symmetry}

Assuming that $\Z_2^\rF \not\subset \U(1)$, 
{which means the spacetime-internal symmetry group structure is
$\Spin \times \U(1)$ structure},
the 6d anomaly polynomial $I_6$ above is in general a linear combination (over $\Z$, if the charge 
$q$ is an integer) of the following two terms:
\begin{equation}
\label{eq:SpinU1-polynomial}
   I^{A}\coloneqq \frac{c_1^3}{6}-\frac{c_1p_1}{24} \in \Z,\qquad I^{B}\coloneqq{c_1^3} \in \Z.
\end{equation}
For a general charge $q$ left-handed Weyl fermion, we have:
\begin{equation}
    I_6=
    q I^A
    +\frac{{q}^3-q}{6} I^B \in \Z.
\end{equation}
$I^A$ and $I^B$ serve as the two generators of $\Hom(\Omega_6^{\Spin}(\B\U(1)),\Z)\cong \Z\times \Z$ by considering their integrals over 6-manifolds representing the elements in the bordism group.
The integer values follow the Atiyah-Singer index theorem for the Dirac operator.
Note that $(q^3-q)/6\in \Z$ for any $q\in \Z$. Instead of $(I^A,I^B)$ as above, we can consider another pair related to it by a $\GL(2,\Z)$ transformation.

\subsection{${\Spin \times_{\Z_2^\rF} \U(1)^\rF}=\Spin^c$ symmetry}

Assuming 
$\Z_2^\rF\subset \U(1)$, 
{which means the spacetime-internal symmetry structure is
$\Spin \times_{\Z_2^\rF} \U(1) \equiv \Spin^c$ structure}
(so that in particular $q$ is necessarily odd for fermions),
the general 6d anomaly polynomial is an integral  linear combination of the following two terms:
\begin{equation}
   \label{spinc-anomaly-basis} 
   I^C\coloneqq \frac{c_1^3}{6}-\frac{c_1p_1}{24}=\frac{(2c_1)^3-(2c_1)p_1}{48} \in \Z,\qquad 
   {I^D\coloneqq {4c_1^3}=\frac{(2c_1)^3}{2}} \in \Z.
\end{equation}
For a general charge $q$ left-handed Weyl fermion, we have:
\begin{equation}
    I_6=
    q I^C
    +\frac{{q}^3-q}{24} I^D \in \Z.
\end{equation}
$I^C$ and $I^D$ serve as the two generators of $\Hom(\Omega_6^{\Spin^c},\Z)\cong \Z\times \Z$,
whose integer values 
following the Atiyah-Singer index theorem for the Dirac operator.

 Note that in this $\Spin^c$ case, 
 $c_1$ is in general not a well-defined integer cohomology class, only $2c_1$ is. This is because in general there is no globally well-defined $\U(1)$ bundle, only $\U(1)/\Z_2$ bundle, the first Chern class of which is {$c_1'=2c_1\in\Z$}.\footnote{{For $\Spin^c$, 
 the $\U(1) \supset {\Z_2^\rF}$ contains the fermion parity as a normal subgroup.\\
$\bullet$ For the original $\U(1)$ with $c_1(\U(1))$, 
the gauge bundle constraint is $w_2(TM)= 2 c_1 \mod 2$.
In the original $\U(1)$, fermions have odd charges under $\U(1)$,
while bosons have even charges under $\U(1)$.
Call the original U(1) gauge field $A$,
then $c_1=\frac{\dd A}{2 \pi} \in \frac{1}{2}\Z$.\\
$\bullet$ For the new $\U(1)'=\frac{\U(1)}{\Z_2^\rF}$ with $c_1(\U(1)')$,
the gauge bundle constraint is $w_2(TM)= c_1' = 2 c_1 \mod 2$.
Call the new $\U(1)'$ gauge field $A'$,
then $c_1'=\frac{\dd A'}{2 \pi}=\frac{\dd (2A)}{2 \pi} = 2 c_1\in 2\frac{1}{2}\Z = \Z$.\\
$\bullet$ To explain why $A' = 2 A$ or $ c_1' = 2 c_1$, we look at the Wilson line operator
$\text{$\exp(\ii q' \oint A')$ and $\exp(\ii q \oint A)$.}$
The original $\U(1)$ has charge transformation $\exp(\ii q \theta)$ with $\theta \in [0, 2 \pi)$,
while the new $\U(1)'$ has charge transformation $\exp(\ii q' \theta')$ with $\theta' \in [0, 2 \pi)$.
But the $\U(1)'=\frac{\U(1)}{\Z_2^\rF}$, so the $\theta=\pi$ in the old $\U(1)$ 
is identified as $\theta'=2\pi$ as a trivial zero
in the new $\U(1)'$.
In the original $\U(1)$, the $q \in \Z$ to be compatible with $\theta \in [0, 2 \pi)$.
In the new $\U(1)'$, the original $q$ is still allowed to have $2\Z$ to be compatible with $\theta \in [0, \pi)$;
but the new $q'=\frac{1}{2} q \in \Z$
and the new $\theta'= 2 \theta  \in [0, 2 \pi)$ are scaled accordingly.
Since the new $q'=\frac{1}{2} q \in \Z$, we show the new $A'=2 A$.}} 

\subsection{${\Spin \times_{\Z_2^\rF} \Z_4^\rF}$ symmetry}

Consider ${\Spin \times_{\Z_2^\rF} \Z_4}$ symmetry ($\subset \Spin^c$ symmetry),
for an odd charge $q \in \Z_4$ Weyl fermion theory in 4d,
we like to match its 4d anomaly to a 5d bordism group index
$\Omega_5^{\Spin \times_{\Z_2^\rF} \Z_4} =\Z_{16}$
or precisely
a 5d cobordism group index
$\Hom(\Omega_5^{\Spin \times_{\Z_2^\rF} \Z_4},\U(1))=\Z_{16}$.

To derive the anomaly index formula
$\upnu(q)\in \Z_{16}$, 
here are some constraints,
\begin{itemize}

\item
For charge-$q=1 \in \Z_4$ Weyl fermion, we have an anomaly index:
$\upnu  =1 \mod 16$, following the Atiyah-Patodi-Singer eta invariant.

\item
For charge-$q=3 = -1 \in \Z_4$ Weyl fermion, we have an anomaly index:
$\upnu  = -1  = 15 \mod 16$, due to the additivity structure of the group of the anomaly index.

\item 
There is a linear map between
the 2-dimensional integral lattice generated by 
two generators $(I^C, I^D)$  of 
$\Hom(\Omega_6^{\Spin^c},\Z)\cong \Z\times \Z$
and the 1-dimensional integral mod 16 lattice
$\Hom(\Omega_5^{\Spin \times_{\Z_2^\rF} \Z_4},\U(1))=\Z_{16}$.
Under that linear map, 
denote that $I^C$ is mapped to $\iota^C \in \Z_{16}$
and $I^D$ is mapped to $\iota^D \in \Z_{16}$.
Then we can solve $\iota^C, \iota^D \in \Z_{16}$
by plugging the constraint of the 
anomaly index $\upnu(q) \in \Z_{16}$ with an odd $q = 1, 3 \in \Z_4$;
we have
\bea
\upnu (q)  &=& q\iota^C  
+\frac{{q}^3-q}{24} \iota^D\mod 16.\cr
\upnu(q=1) &=& 1 = 1 \iota^C + 0  \iota^D \mod 16.\cr
\upnu(q=3) &=& -1 = 3 \iota^C + 1  \iota^D \mod 16 . \cr
 &\Rightarrow&  \iota^C =1 \mod 16, \quad \iota^D =-4 \mod 16.
\eea
\end{itemize}
Finally, 
for an odd charge $q \in \Z_4$ Weyl fermion, 
we obtain its anomaly index formula
\bea
\upnu (q)  = q\iota^C  
+\frac{{q}^3-q}{24} \iota^D
= q - 4 \frac{{q}^3-q}{24} =  
\frac{- q^3+ 7q}{6}
\mod 16 \in \Z_{16}.
\eea

Using the theory in \cite{Cheng:2024awi2411.05786},
we find that the $\Z_{16}$ class of 
5d $\Z_{4}^\rF$ fSPTs and $C_{2} \times \Z_2^\rF$ fSPTs
has: 
\begin{itemize}

\item
For $\nu=N=2 \text{ or even } \in \Z_{16}$, we have a symmetric gapped $\Z_4$ gauge theory TQFT.

\item
For $\nu=N/2=1 \text{ or odd }\in \Z_{16}$,
we find a non-TQFT symmetric gapped state via stacking lower-dimensional (2+1)d non-discrete-gauge-theory topological order (that has TQFT descriptions) inhomogeneously.

\end{itemize}

There are two applications for $\Z_4^\rF$ symmetry:
\begin{enumerate}
\item First application: The $\Z_{4,X}$ symmetry as $\Z_4^\rF$.    
There are 3 missing right-handed neutrinos not found in the Standard Model, 
each of such $\bar{\nu}_R$ 
has $\Z_{4}^\rF$ charge $1$,
thus for 3 families of such $\bar{\nu}_R$, 
we have the total anomaly index
\bea
3(1 \mod 16) =(3 \mod 16).
\eea
The complementary anomaly index of the 3-family SM 
(without 3 $\bar{\nu}_R$) is 
\bea
(-3 \mod 16).
\eea
Using a $\nu=\text{even} \in \Z_{16}$ index 
symmetric gapped $\Z_4$ gauge theory TQFT \cite{Cheng:2024awi2411.05786} is not enough to cancel the SM's $N_f=3$ anomaly, either some $\nu_R$, or some 4d gapless vector, or some extra dimensional 5d bulk, or some non-TQFT kind of fracton topological order is needed.
\item Second application: The $\Z_{2 N_f=4, {{ \mathbf{B} + \mathbf{L}}}}$ symmetry as $\Z_4^\rF$
for $N_f=2$.    
There are 2 missing right-handed neutrinos not found in the $N_f=2$-family Standard Model, 
each of such $\bar{\nu}_R$ 
has $\Z_{4}^\rF$ charge $-1$,
thus for 3 families of such $\bar{\nu}_R$, 
we have the total anomaly index
\bea
2(-1 \mod 16) =(-2 \mod 16).
\eea
The complementary anomaly index of the $N_f=2$-family SM 
(without 2 $\bar{\nu}_R$) is 
\bea \label{eq:Spin-2Nf=4}
(2 \mod 16).
\eea
Note that the $2 \mod 16$ corresponds to a 5d fermionic SPT which is \emph{beyond the bosonic group cohomology} description.
Using the theory in \cite{Cheng:2024awi2411.05786},
for $\nu=N=2 \in \Z_{16}$, we have a symmetric gapped fermionic $\Z_4$ gauge theory TQFT.
\end{enumerate}
In a one lower dimensional analogy,
assume there is a 3d {$\Pin^+$} TQFT with anomaly described by the 4d effective action 
{$S_{4d}=-\upnu (2\pi \eta/16)$}
with{an anomaly index} $-\upnu\in \Z_{16}=\Hom(\Omega^{{\Pin^+}}_{{4}},\U(1))$, such TQFTs were considered in 
\cite{Fidkowski1305.5851:2013jua, 
Metlitski1406.3032:2014xqa,
Wang1610.04624:2016qkb,
Tachikawa:2016cha,
Tachikawa:2016nmo,cheng2018microscopic1707.02079, Tata2104.14567:2021jwp}.

\subsection{${\Spin \times_{\Z_2^\rF} \Z_8^\rF}$ symmetry}

Consider ${\Spin \times_{\Z_2^\rF} \Z_8}$ symmetry ($\subset \Spin^c$ symmetry),
for an odd charge $q \in \Z_8$ Weyl fermion theory in 4d,
we like to match its 4d anomaly to a 5d bordism group index
$\Omega_5^{\Spin \times_{\Z_2^\rF} \Z_8}=\Z_{32} \times \Z_2$
or precisely
a 5d cobordism group index
$\Hom(\Omega_5^{\Spin \times_{\Z_2^\rF} \Z_8},\U(1))=\Z_{32} \times \Z_2$.

To derive the anomaly index formula
$(\upnu_1(q), \upnu_2(q)) \in \Z_{32} \times \Z_2$, here are some constraints,
\begin{itemize}
\item
For charge-$q=1 \in \Z_8$ Weyl fermion, we have an anomaly index:
$\upnu_1  =1 \mod 32$, following the Atiyah-Patodi-Singer eta invariant.
\item
For charge-$q=7 = -1 \in \Z_8$ Weyl fermion, we have an anomaly index:
$\upnu_1  = -1  = 31 \mod 32$, due to the additivity structure of the group of the anomaly index.
\item 
There is a linear map between
the 2-dimensional integral lattice generated by 
two generators $(I^C, I^D)$  of 
$\Hom(\Omega_6^{\Spin^c},\Z)\cong \Z\times \Z$
and the 2-dimensional integral mod $(32,2)$ lattice
$\Hom(\Omega_5^{\Spin \times_{\Z_2^\rF} \Z_8},\U(1))=\Z_{32} \times \Z_2$.
Under that linear map, 
denote that 
$I^C$ is mapped to $(\iota^C_1, \iota^C_2) \in \Z_{32} \times \Z_2$
and $I^D$ is mapped to 
$(\iota^D_1, \iota^D_2) \in \Z_{32} \times \Z_2$.
Then we can solve $\iota^C_1, \iota^D_1 \in \Z_{32}$
by plugging the constraint of the 
anomaly index $\upnu_1(q) \in \Z_{32}$ with an odd $q = 1, 7 \in \Z_8$;
we have
\bea
\upnu_1(q) &=& q\iota^C_1  
+\frac{{q}^3-q}{24} \iota^D_1 \mod 32.\cr
\upnu_1(q=1) &=& 1 = 1 \iota^C_1 + 0  \iota^D_1 \mod 32.\cr
\upnu_1(q=7) &=& -1 = 7 \iota^C_1 + 14  \iota^D_1 \mod 32 . \cr
 &\Rightarrow&  \iota^C_1 =1 \mod 32, \quad \iota^D_1 =4 \mod 32.
\eea
Plugging the constraint of the 
anomaly index $\upnu_2(q) \in \Z_{2}$ with an odd $q = 1, 7 \in \Z_8$;
we have
\bea
\upnu_2(q)  &=&  q\iota^C_2  
+\frac{{q}^3-q}{24} \iota^D_2 \mod 2.\cr
\upnu_2(q=1) &=& 0 = 1 \iota^C_2 + 0  \iota^D_2 \mod 2.\cr
\upnu_2(q=7) &=& 0 = 7 \iota^C_2 + 14  \iota^D_2 \mod 2 . \cr
 &\Rightarrow&  \iota^C_2 =0 \mod 2, \quad \iota^D_2 =0, \text{ or } 1 \mod 2.
\eea
Note that $\iota^D_2=0$ or $1$ cannot be solved directly here,
but we expect that a nontrivial map thus $\iota^D_2=1$  which can be verified in the next step. 
\item Via Ref.~\cite{Hsieh2018ifc1808.02881} anomaly index formula (2.49),
we are able to derive an appropriate linear combination of two generators in (2.49) to give rise to
\bea
(\upnu_1(q=3),\upnu_2(q=3)) &=& ( 7 \mod 32 , \quad 1 \mod 2) .\cr
(\upnu_1(q=5),\upnu_2(q=5)) &=& ( -7 \mod 32 , \quad 1 \mod 2) .\cr
\upnu_1(q=3) &=& 7 = 3 \iota^C_1 + 1  \iota^D_1 \mod 32.\cr
\upnu_1(q=5) &=& -7 = 5 \iota^C_1 + 5  \iota^D_1 \mod 32 . \cr
 &\Rightarrow&  \iota^C_1 =1 \mod 32, \quad \iota^D_1 =4 \mod 32.\cr
 \upnu_2(q=3) &=& 1 = 3 \iota^C_2 + 1  \iota^D_2 \mod 2.\cr
\upnu_2(q=5) &=& 1 = 5 \iota^C_2 + 5  \iota^D_2 \mod 2 . \cr
 &\Rightarrow&  \iota^C_2 =0 \mod 2, \quad \iota^D_2 = 1 \mod 2.
\eea
\end{itemize}
Thus
for an odd charge $q \in \Z_8$ Weyl fermion, 
we obtain its anomaly index formula
\bea 
&&(\upnu_1(q), \upnu_2(q)) \in \Z_{32} \times \Z_2 \nn \\
&=& (q\iota^C_1  
+\frac{{q}^3-q}{24} \iota^D_1 \mod 32, 
q\iota^C_2  
+\frac{{q}^3-q}{24} \iota^D_2 \mod 2)   \nn \\
&=& (\frac{q^3 + 5 q}{6} \mod 32,  \frac{q^3 - q}{24} \mod 2). 
\eea
For a charge $q=1,3,5,7$ Weyl fermion, 
we have a map to the anomaly index
\bea
q=1,3,5,7 \mapsto 
(\upnu_1, \upnu_2) =(1, 0), (7, 1), (-7, 1), (-1, 0).
\eea

Using the theory in \cite{Cheng:2024awi2411.05786},
we find that the $\Z_{32}$ class of 
$\Z_{8}^\rF$ fSPTs and $C_{4} \times \Z_2^\rF$ fSPTs
has: 
\begin{itemize}

\item
For $\nu=N=4 \in \Z_{32}$, we have a symmetric gapped $\Z_4$ gauge theory TQFT.

\item
For $\nu=N/2=2 \in \Z_{32}$,
we find a non-TQFT symmetric gapped state via stacking lower-dimensional (2+1)d non-discrete-gauge-theory topological order (that has TQFT descriptions) inhomogeneously.

\item
For $\nu=1 \in \Z_{32}$, we do not have either of
symmetric gapped states.

\end{itemize}

We consider two applications for $\Z_8^\rF$ symmetry:
\begin{enumerate}
    
\item First application: The $\Z_{8,X}$ symmetry as $\Z_8^\rF$.
Missing 3 right-handed neutrinos (not found in the SM), each has $\Z_{8,X}$ charge $q=5$, thus the 3 right-handed neutrinos has 
\bea
3(-7 \mod 32, \quad 1 \mod 2) =(11 \mod 32, \quad 1 \mod 2).
\eea
The complementary anomaly index of the SM is 
\bea
(21 \mod 32, \quad 1 \mod 2).
\eea

\item Second application: The $\Z_{2N_f=8, {{ \mathbf{B} + \mathbf{L}}}}^\rF$ symmetry as $\Z_8^\rF$ with $N_f=4$.
Missing right-handed neutrinos (not found in the SM), each has 
$\Z_{2N_f=8, {{ \mathbf{B} + \mathbf{L}}}}^\rF$ charge $q=-1=7 \mod 8$, thus the 4 right-handed neutrinos has 
\bea
4(-1 \mod 32, \quad 0 \mod 2) =(-4 \mod 32, \quad 0 \mod 2).
\eea
The complementary anomaly index of the SM is 
\bea
(4 \mod 32, \quad 0 \mod 2),
\eea
thus anomalous. Note that the $4 \mod 32$ class of 
5d $\Z_8^\rF$ fermionic SPTs is still a beyond group cohomology SPTs. 
(The situation is different for $N_f=3$
which gives rise to a group cohomology SPTs, shown in \Sec{sec:Z6F}
and \Sec{sec:Z9-fSPT}.)

\end{enumerate}

\subsection{${\Spin \times \Z_3}$ symmetry}

\label{sec:Z3}

Consider ${\Spin \times  \Z_3}$ symmetry ($\subset \Spin \times \U(1)$ symmetry),
for an integer charge $q \in \Z_3$ Weyl fermion theory in 4d,
we like to match its 4d anomaly to a 5d bordism group index
$\Omega_5^{\Spin \times  \Z_3} =\Z_{9}$
or precisely
a 5d cobordism group index
$\Hom(\Omega_5^{\Spin \times  \Z_3},\U(1))=\Z_{9}$.

To derive the anomaly index formula
$\upnu(q)\in \Z_{9}$, 
here are some constraints,
\begin{itemize}

\item
For charge-$q=1 \in \Z_3$ Weyl fermion, we have an anomaly index:
$\upnu  =1 \mod 16$, following the Atiyah-Patodi-Singer eta invariant.

\item
For charge-$q=2 = -1 \in \Z_3$ Weyl fermion, we have an anomaly index:
$\upnu  = -1  = 8 \mod 9$, due to the additivity structure of the group of the anomaly index.

\item 
There is a linear map between
the 2-dimensional integral lattice generated by 
two generators $(I^A, I^B)$  of 
$\Hom(\Omega_6^{\Spin \times \U(1)},\Z)\cong \Z\times \Z$
and the 1-dimensional integral mod 9 lattice
$\Hom(\Omega_5^{\Spin \times \Z_3},\U(1))=\Z_{9}$.
Under that linear map, 
denote that $I^A$ is mapped to $\iota^A \in \Z_{9}$
and $I^B$ is mapped to $\iota^B \in \Z_{9}$.
Then we can solve $\iota^A, \iota^B \in \Z_{9}$
by plugging the constraint of the 
anomaly index $\upnu(q) \in \Z_{9}$ with  $q = 1, 2 \in \Z_3$;
we have
\bea
\upnu (q)  &=& q\iota^A  
+\frac{{q}^3-q}{6} \iota^B\mod 9.\cr
\upnu(q=1) &=& 1 = 1 \iota^A + 0  \iota^B \mod 9.\cr
\upnu(q=2) &=& -1 = 2 \iota^A + 1  \iota^B \mod 9 . \cr
 &\Rightarrow&  \iota^A =1 \mod 9, \quad \iota^B =-3 \mod 9.
\eea
\end{itemize}
Finally, for a charge $q \in \Z_3$ Weyl fermion, 
we obtain its anomaly index formula
\bea
\upnu (q)  = q\iota^A  
+\frac{{q}^3-q}{6} \iota^B
 =\frac{- q^3+ 3q}2
\mod 9 \in \Z_{9}.
\eea
For a charge $q=1,2$ Weyl fermion, 
we have a map to the anomaly index
\bea
q=1,2 \mapsto 
\upnu = 1, -1 \mod 9  \in \Z_{9}.
\eea
We consider two applications for $\Z_2^\rF \times \Z_3$ symmetry:
\begin{enumerate}
\item First application: The $\Z_{2N_f=6, {{ \mathbf{B} + \mathbf{L}}}}^\rF$ symmetry as $\Z_6^\rF$ with $N_f=3$.

There are 3 missing right-handed neutrinos not found in the Standard Model, 
each of such $\bar{\nu}_R$ 
has $\Z_{2N_f=6}^\rF$ charge $-1=5$, 
each has $\Z_{3}$ charge $2$,\footnote{We
can label $n_6 \in \Z_6 = \Z_6^\rF \supset  \Z_2^\rF$ in terms of a doublet $(n_2^\rF, n_3) \in  
\Z_2^\rF \times \Z_3$,
such that the bosons have $n_2^\rF=0$
and the fermions have $n_2^\rF=1$. 
In addition, without loss of generality, 
we assign the charge $q=1 \in \Z_6^\rF$ fermion
to the $(n_2^\rF, n_3)=(1,1) \in  \Z_2^\rF \times \Z_3 $. These are enough to constrain
the map in between as $n_6 = 3 n_2^\rF - 2 n_3$. Thus, $n_6=0,1,2,3,4,5$ is mapped to
$(n_2^\rF, n_3)=(0,0), (1,1), (0,2),
(1,0), (0,1), (1,2)$.
} 
thus for 3 families of such $\bar{\nu}_R$, 
we have the total anomaly index
\bea
3(-1 \mod 9) =(6 \mod 9).
\eea
The complementary anomaly index of the 3-family SM is 
\bea
(3 \mod 9).
\eea
This result shows that $N_f=3$ is special
which gives rise to a nontrivial
$\Z_3$ class of group cohomology SPTs, see \Sec{sec:Z9-fSPT}.
\item
Second application: The $\Z_{2 N_c N_f=6, {{ \mathbf{Q} + N_c \mathbf{L}}}}$ symmetry as $\Z_6^\rF$
for $N_c=3$ and $N_f=1$.    
There are 1 missing right-handed neutrinos not found in the $N_f=1$-family Standard Model, 
each of such $\bar{\nu}_R$ 
has $\Z_{6}^\rF$ charge $-3=3$ that has 
$\Z_3$ charge 0,
thus for 1 family of such $\bar{\nu}_R$, 
we have the total anomaly index
\bea
1(0 \mod 9) =(0 \mod 9).
\eea
The complementary anomaly index of the $N_f=1$-family SM 
(without a $\bar{\nu}_R$) is 
\bea
(0 \mod 9),
\eea
still anomaly-free.
\end{enumerate}

\subsection{${\Spin \times_{\Z_2^\rF} \Z_{6}^\rF}$ symmetry}
\label{sec:Z6F}

Consider ${\Spin \times_{\Z_2^\rF} \Z_6^\rF}$ symmetry ($\subset \Spin^c$ symmetry),
for an odd charge $q \in \Z_6^\rF$ Weyl fermion theory in 4d,
we like to match its 4d anomaly to a 5d bordism group index
$\Omega_5^{\Spin \times_{\Z_2^\rF} \Z_6^\rF} =
\Omega_5^{\Spin \times \Z_3} =
\Z_{9}$
or precisely
a 5d cobordism group index
$\Hom(\Omega_5^{\Spin \times_{\Z_2^\rF} \Z_6^\rF},\U(1))=
\Hom(\Omega_5^{\Spin \times \Z_3},\U(1))
= \Z_{9}$.

To derive the anomaly index formula
$\upnu(q) \in \Z_{9}$, here are some constraints,
\begin{itemize}
\item
For charge-$q=1 \in \Z_6$ Weyl fermion, we have an anomaly index:
$\upnu   =1 \mod 9$, following the Atiyah-Patodi-Singer eta invariant.
\item
For charge-$q=5 = -1 \in \Z_6$ Weyl fermion, we have an anomaly index:
$\upnu   = -1  = 8 \mod 9$, due to the additivity structure of the group of the anomaly index.

\item 
There is a linear map between
the 2-dimensional integral lattice generated by 
two generators $(I^C, I^D)$  of 
$\Hom(\Omega_6^{\Spin^c},\Z)\cong \Z\times \Z$
and the 1-dimensional integral mod 9 lattice
$\Hom(\Omega_5^{\Spin \times_{\Z_2^\rF} \Z_6},\U(1))=\Z_{9}$.
Under that linear map, 
denote that $I^C$ is mapped to $\iota^C \in \Z_{9}$
and $I^D$ is mapped to $\iota^D \in \Z_{9}$.
Then we can solve $\iota^C, \iota^D \in \Z_{9}$
by plugging the constraint of the 
anomaly index $\upnu(q) \in \Z_{16}$ with an odd $q = 1, 5 \in \Z_6$;
we have
\bea
\upnu (q)  &=& q\iota^C  
+\frac{{q}^3-q}{24} \iota^D\mod 9.\cr
\upnu(q=1) &=& 1 = 1 \iota^C + 0  \iota^D \mod 9.\cr
\upnu(q=5) &=& -1 = 5 \iota^C + 5  \iota^D \mod 9 . \cr
 &\Rightarrow&  \iota^C =1 \mod 9, \quad \iota^D =-3 \mod 9.
\eea
\end{itemize}
Finally, 
for an odd charge $q \in \Z_6$ Weyl fermion, 
we obtain its anomaly index formula
\bea
\upnu (q) =
q\iota^C  
+\frac{{q}^3-q}{24} \iota^D
=
\frac{-{q}^3+ 9 q}{8} 
\mod 9 \in \Z_{9}
\eea
For a charge $q=1,3,5$ Weyl fermion, 
we have a map to the anomaly index
\bea
q=1,3,5 \mapsto 
\upnu = 1, 0, -1 \mod 9  \in \Z_{9}.
\eea

We consider two applications for $\Z_6^\rF$ symmetry:
\begin{enumerate}
\item First application: The $\Z_{2N_f=6, {{ \mathbf{B} + \mathbf{L}}}}^\rF$ symmetry as $\Z_6^\rF$ with $N_f=3$.
There are 3 missing right-handed neutrinos not found in the Standard Model, 
each of such $\bar{\nu}_R$ 
has $\Z_{2N_f=6, {{ \mathbf{B} + \mathbf{L}}}}^\rF$ charge $-1=5$,
thus for 3 families of such $\bar{\nu}_R$, 
we have the total anomaly index
\bea
3(-1 \mod 9) =(6 \mod 9).
\eea
The complementary anomaly index of the 3-family SM 
(without 3 $\bar{\nu}_R$) is 
\bea
(3 \mod 9).
\eea
This result shows that $N_f=3$ is special
which gives rise to a nontrivial
$\Z_3$ class of group cohomology SPTs, see \Sec{sec:Z9-fSPT}.

\item Second application: The $\Z_{2 N_c N_f=6, {{ \mathbf{Q} + N_c \mathbf{L}}}}$ symmetry as $\Z_6^\rF$
for $N_c=3$ and $N_f=1$.    
There are 1 missing right-handed neutrinos not found in the $N_f=1$-family Standard Model, 
each of such $\bar{\nu}_R$ 
has $\Z_{6}^\rF$ charge $-3=3$,
thus for 1 family of such $\bar{\nu}_R$, 
we have the total anomaly index
\bea
1(0 \mod 9) =(0 \mod 9).
\eea
The complementary anomaly index of the $N_f=1$-family SM 
(without a $\bar{\nu}_R$) is 
\bea
(0 \mod 9),
\eea
still anomaly-free.

\end{enumerate}

\subsection{${\Spin \times_{\Z_2^\rF} \Z_{12}^\rF}$ symmetry}

Consider ${\Spin \times_{\Z_2^\rF} \Z_{12}^\rF}$ symmetry ($\subset \Spin^c$ symmetry),
for an odd charge $q \in \Z_{12}^\rF$ Weyl fermion theory in 4d,
we like to match its 4d anomaly to a 5d bordism group index
$\Omega_5^{\Spin \times_{\Z_2^\rF} \Z_{12}^\rF} =
\Omega_5^{\Spin \times_{\Z_2^\rF} \Z_{4}^\rF \times \Z_3} =
\Z_{16} \times
\Z_{9}$
or precisely
a 5d cobordism group index
$\Hom(\Omega_5^{\Spin \times_{\Z_2^\rF} \Z_{12}^\rF},\U(1))=
\Hom(\Omega_5^{\Spin \times_{\Z_2^\rF} \Z_{4}^\rF \times \Z_3},\U(1))
= \Z_{16} \times
\Z_{9}$.

To derive the anomaly index formula
$(\upnu_1(q),\upnu_2(q))  \in \Z_{16} \times \Z_9$, 
we follow the previous arguments with similar constraints. 

For an odd charge $q \in \Z_{12}$ Weyl fermion, 
We arrive at the map to the anomaly index
\bea \label{eq:Spin-Z12}
&& q=1, 3, 5, 7, 9, 11 \cr 
&&\mapsto 
(\upnu_1, \upnu_2) =
(1, 1), (-1, 0), (1, -1), (-1, 1), (1, 0), (-1, -1) 
\in \Z_{16} \times \Z_9.
\eea

We provide two applications for $\Z_{12}^\rF$ symmetry:
\begin{enumerate}
\item First application: The $\Z_{12,X}$ symmetry as $\Z_{12}^\rF$.    
There are 3 missing right-handed neutrinos not found in the Standard Model, 
each of such $\bar{\nu}_R$ 
has $\Z_{12}^\rF$ charge $5$,
thus for 3 families of such $\bar{\nu}_R$, 
we have the total anomaly index
\bea
3(1 \mod 16, -1 \mod 9) =(3 \mod 16, -3 \mod 9).
\eea
The complementary anomaly index of the 3-family SM 
(without 3 $\bar{\nu}_R$) is 
\bea
(-3 \mod 16, 3 \mod 9)).
\eea
\item Second application:
The $\Z_{2 N_c N_f=12, {{ \mathbf{Q} + N_c \mathbf{L}}}}$ symmetry as $\Z_{12}^\rF$
for $N_c=3$ and $N_f=2$.    
Suppose there are 2 missing right-handed neutrinos not found in the 
$N_f=2$ Standard Model, 
each such $\bar{\nu}_R$ 
has $\Z_{2N_c N_f= 12}^\rF$ charge $-3=9$, 
thus for 2 families of such $\bar{\nu}_R$, 
we have 
\bea
2(1 \mod 16, \quad 0 \mod 9) =(2 \mod 16, \quad 0 \mod 9).
\eea
The complementary anomaly index of the 2-family SM 
(without 2 $\bar{\nu}_R$) is 
\bea
(-2 \mod 16, \quad 0 \mod 9),
\eea
thus anomalous.
This result, up to the orientation of the charge, matches with \eq{eq:Spin-2Nf=4}, between the 
$\Z_{2 N_f=4, {{ \mathbf{B} + \mathbf{L}}}}$
there and 
$\Z_{2 N_c N_f=12, {{ \mathbf{Q} + N_c \mathbf{L}}}}$ 
here for $N_f=2$.\footnote{We
can label $n_{12} \in \Z_{12} = \Z_{12}^\rF \supset  \Z_2^\rF$ in terms of a doublet $(n_4^\rF, n_3) \in  
\Z_4^\rF \times \Z_3$,
such that the bosons have $n_4^\rF=0,2$
and the fermions have $n_2^\rF=1,3$. 
In addition, without loss of generality, 
we assign the charge $q=1 \in \Z_{12}^\rF$ fermion
to the $(n_4^\rF, n_3)=(1,1) \in  \Z_4^\rF \times \Z_3 $. 
These are enough to constrain
the map in between as $n_{12} = - 3 n_4^\rF + 4 n_3$. Thus, 
$n_{12}=0,2,4,6,8,10$ is mapped to
$(n_4^\rF, n_3)=(0,0), (2,2), (0,1),
(2,0), (0,2), (2,1)$
while
$n_{12}=1,3,5,7,9,11$ is mapped to
$(n_4^\rF, n_3)=(1,1), (3,0), (1,2),
(3,1), (1,0), (3,2)$.
} 
\end{enumerate}
 
\newpage

\subsection{${\Spin \times \Z_9}$ symmetry}

Consider ${\Spin \times  \Z_9}$ symmetry ($\subset \Spin \times \U(1)$ symmetry),
for an integer charge $q \in \Z_9$ Weyl fermion theory in 4d,
we like to match its 4d anomaly to a 5d bordism group index
$\Omega_5^{\Spin \times  \Z_9} =
\Z_{27} \times \Z_3$
or precisely
a 5d cobordism group index
$\Hom(\Omega_5^{\Spin \times  \Z_9},\U(1))=\Z_{27} \times \Z_3$.
\begin{itemize}
\item
For charge-$q=1 \in \Z_9$ Weyl fermion, we have an anomaly index:
$\upnu_1  =1 \mod 27$, following the Atiyah-Patodi-Singer eta invariant.
\item
For charge-$q=8 = -1 \in \Z_9$ Weyl fermion, we have an anomaly index:
$\upnu_1  = -1  = 26 \mod 27$, due to the additivity structure of the group of the anomaly index.
\item 
There is a linear map between
the 2-dimensional integral lattice generated by 
two generators $(I^A, I^B)$  of 
$\Hom(\Omega_6^{\Spin \times \U(1)},\Z)\cong \Z\times \Z$
and the 2-dimensional integral mod $(27,3)$ lattice
$\Hom(\Omega_5^{\Spin \times \Z_9},\U(1))=\Z_{27} \times \Z_3$.
Under that linear map, 
denote that 
$I^A$ is mapped to $(\iota^A_1, \iota^A_2) \in \Z_{27} \times \Z_3$
and $I^B$ is mapped to 
$(\iota^B_1, \iota^B_2) \in \Z_{27} \times \Z_3$.
Then we can solve $\iota^A_1, \iota^B_1 \in \Z_{27}$
by plugging the constraint of the 
anomaly index $\upnu_1(q) \in \Z_{27}$ with $q = 1, 8 \in \Z_9$;
we have
\bea
\upnu_1(q) &=& q\iota^A_1  
+\frac{{q}^3-q}{6} \iota^B_1 \mod 27.\cr
\upnu_1(q=1) &=& 1 = 1 \iota^A_1 + 0  \iota^B_1 \mod 27.\cr
\upnu_1(q=8) &=& -1 = 8 \iota^A_1 + 84  \iota^B_1 \mod 27 . \cr
 &\Rightarrow&  \iota^A_1 =1 \mod 27, \quad \iota^B_1 = 6 \mod 27.
\eea
Plugging the constraint of the 
anomaly index $\upnu_2(q) \in \Z_{3}$ with $q = 1, 8 \in \Z_9$;
we have
\bea
\upnu_2(q)  &=&  q\iota^A_2  
+\frac{{q}^3-q}{6} \iota^B_2 \mod 3.\cr
\upnu_2(q=1) &=& 0 = 1 \iota^A_2 + 0  \iota^B_2 \mod 3.\cr
\upnu_2(q=8) &=& 0 = 8 \iota^A_2 + 84  \iota^B_2 \mod 3 . \cr
 &\Rightarrow&  \iota^A_2 =0 \mod 3, \quad \iota^B_2 =0,  1, \text{ or } 2 \mod 3.
\eea
Note that $\iota^B_2=0,1,2$ 
cannot be solved directly here,
but we expect that a nontrivial map thus $\iota^B_2=1$ or $2$    which can be verified in the next step. 
\item Via Ref.~\cite{Hsieh2018ifc1808.02881} anomaly index formula (2.31),
we are able to derive an appropriate linear combination of two generators in (2.31) to give rise to
\bea
(\upnu_1(q=3),\upnu_2(q=3)) &=& ( 0 \mod 27 , \quad 2 \mod 3) .\cr
 \upnu_2(q=3) &=& 2 = 3 \iota^A_2 + 4  \iota^B_2 \mod 3.\cr
 &\Rightarrow&  \iota^A_2 =0 \mod 3, \quad \iota^B_2 = 2 \mod 3.
\eea
\end{itemize}
Thus
for a charge $q \in \Z_9$ Weyl fermion, 
we obtain its anomaly index formula
\bea 
&&(\upnu_1(q), \upnu_2(q)) \in \Z_{27} \times \Z_3 \nn \\
&=& (
 q\iota^A_1  
+\frac{{q}^3-q}{6} \iota^B_1 \mod 27
, 
q\iota^A_2  
+\frac{{q}^3-q}{6} \iota^B_2 \mod 3)   \nn \\
&=& (q^3 \mod 27,  \frac{q^3 - q}{3} \mod 3). 
\eea
For a charge $q \in \Z_9$ Weyl fermion, 
we have a map to the anomaly index
\bea
q=1,2,3,4,5,6,7,8 \mapsto 
(\upnu_1, \upnu_2) 
 = (1, 0), (8, 2),  (0, 2), (10, 2),  (17, 1), (0, 1),  (19, 1),  (26, 0).
\eea
There are 3 missing right-handed neutrinos not found in the Standard Model, 
each such $\bar{\nu}_R$ 
has $\Z_{2N_c N_f= 18}^\rF$ charge $-3=15$, 
each has $\Z_{9}$ charge $6$,\footnote{We
can label $n_{18} \in \Z_{18} = \Z_{18}^\rF \supset  \Z_2^\rF$ in terms of a doublet 
$(n_2^\rF, n_9) \in \Z_2^\rF \times  \Z_9$,
such that the bosons have $n_2^\rF=0$
and the fermions have $n_2^\rF=1$. 
In addition, without loss of generality, 
we assign the charge $q=1 \in \Z_{18}^\rF$ fermion
to the $( n_2^\rF, n_9)=(1,1) \in \Z_2^\rF \times  \Z_9$. These are enough to constrain
the map in between as $n_{18} = 9 n_2^\rF - 8 n_9$. Thus, $n_{18}=0, 1, 2, 3, 4, 5, 6, 7, 8, 9, 10, 11, 12, 13, 14, 15, 16, 17$ is mapped to
$(n_2^\rF, n_9)=
(0, 0), (1, 1), 
(0, 2), (1, 3), (0, 4), (1, 5), (0, 6), (1, 7), (0, 
8), (1, 0), (0, 1), (1, 2), (0, 3), (1, 4), (0, 5), (1, 6), (0, 7), 
(1, 8)$.
} thus for 3 families of such $\bar{\nu}_R$, 
we have 
\bea
3(0 \mod 27, \quad 1 \mod 3) =(0 \mod 27, \quad 0 \mod 3).
\eea
The complementary anomaly index of the 3-family SM 
(without 3 $\bar{\nu}_R$) is 
\bea
(0 \mod 27, \quad 0 \mod 3),
\eea
thus anomaly-free.

\subsection{${\Spin \times_{\Z_2^\rF} \Z_{18}^\rF}$ symmetry}

Consider ${\Spin \times_{\Z_2^\rF} \Z_{18}^\rF}$ symmetry ($\subset \Spin^c$ symmetry),
for an odd charge $q \in \Z_18^\rF$ Weyl fermion theory in 4d,
we like to match its 4d anomaly to a 5d bordism group index
$\Omega_5^{\Spin \times_{\Z_2^\rF} \Z_{18}^\rF} =
\Omega_5^{\Spin \times \Z_9} =
\Z_{27} \times \Z_3$
or precisely
a 5d cobordism group index
$\Hom(\Omega_5^{\Spin \times_{\Z_2^\rF} \Z_{18}},\U(1))=
\Hom(\Omega_5^{\Spin \times \Z_9},\U(1))
= \Z_{27} \times \Z_3$.

To derive the anomaly index formula
$(\upnu_1(q),\upnu_2(q))  \in \Z_{27} \times \Z_3$, 
we follow the previous arguments with similar constraints. 

For an odd charge $q \in \Z_{18}$ Weyl fermion, 
We arrive at the map to the anomaly index
\bea
&& q=1, 3, 5, 7, 9, 11, 13, 15, 17 \cr 
&&\mapsto 
(\upnu_1, \upnu_2) =
(1, 0), (0, 2), (17, 1), (19, 1), (0, 0), (8, 2), (10, 2), (0, 1), 
(26, 0).
\eea
There are 3 missing right-handed neutrinos not found in the Standard Model, 
each such $\bar{\nu}_R$ 
has $\Z_{2N_c N_f= 18}^\rF$ charge $-3=15$, 
thus for 3 families of such $\bar{\nu}_R$, 
we have 
\bea
3( 0 \mod 27, \quad 1 \mod 3) =(0 \mod 27, \quad 0 \mod 3).
\eea
There is an ambiguity of the first mod 27 index, where the ambiguity comes 
$9k \mod 27$ for $k \in \Z$
thus at $0 \mod 9$;
but the outcome choice would not affect our  $N_f=3$ anomaly,
because $9 N_f = 27 = 0 \mod 27$.
The complementary anomaly index of the 3-family SM 
(without 3 $\bar{\nu}_R$) is 
\bea
(0 \mod 27, \quad 0 \mod 3),
\eea
thus anomaly-free.

\section{$\Z_9$ class 
topological invariants of
$\Spin \times \Z_3$ and the group extension
$1 \to \Z_3 \to \Z_9 \to  \Z_3 \to 1$}
\label{sec:Z9-fSPT}

Here we demonstrate the 
bordism $\Omega_5^{\Spin \times \Z_3}= \Z_9$ or the cobordism
$\Hom(\Omega_5^{\Spin \times \Z_3},\U(1)) = \Z_9$ class 
topological invariants of
$\Spin \times \Z_3$ symmetry; 
such that the magical $\Z_9$ class forms a group extension:
\bea
1 \to \Z_3 \to &\Z_9& \to  \Z_3 \to 1, 
\eea
or more schematically 
\bea \label{eq:Z9class}
1 \to (\Z_3)_{\text{group cohomology}} \to &(\Z_9)_{\text{full cobordism class}}& \to (\Z_3)_{\text{beyond group cohomology}} \to 1.
\eea
The normal $\Z_3$ is from the group cohomology bosonic SPT class in 
$\H^5(\B \Z_3, \U(1))= \Z_3$ tensor product with the free fermion SPT class.
The quotient $\Z_9/\Z_3=\Z_3$ is from the
beyond group cohomology SPT class.

To demonstrate the above statement, 
we shall embed
$\Spin \times \Z_3 \subset \Spin \times \U(1)$ first, then change the U(1) gauge field to a $\Z_3$ gauge field.

We take the $\Spin \times \U(1)$ structure's
6d anomaly polynomial $I_6$ above in
\eq{eq:SpinU1-polynomial} with $q=1$ Weyl fermion gives the following 6d anomaly polynomial's invertible field theory
\begin{equation}
   \exp( \ii k \theta \int_{M^6} \frac{c_1^3}{6}-\frac{c_1p_1}{24})
\end{equation}   
with $k \in \Z$, while $\theta \in [0, 2 \pi)$,
and 
$\int_{M^6} \frac{c_1^3}{6}-\frac{c_1p_1}{24}
 \in \Z$ on a closed $M^6$ of the given structure. The 5d manifold at the interface of jumping $\theta=0$ to $\theta=2 \pi$ gives a 5d invertible field theory (as a 5d SPTs):
\begin{equation}
   \exp( \ii k  \int_{M^5} A \frac{c_1^2}{6}-A \frac{p_1}{24}).
\end{equation} 
Now we redefine the U(1) gauge field $A$ as a $\Z_3$ gauge field $\tilde A \in \Z_3$
with the following replacement:
\bea
A &\mapsto& \frac{2 \pi}{3}  \tilde A.\cr 
c_1 = \frac{\dd A}{2 \pi} &\mapsto& 
\frac{\dd  \tilde A}{3} \equiv \beta_{(3,3) }\tilde A. 
\eea
The $\beta_{(n,m)}: \H^*(-,\Z_m) \mapsto 
\H^{*+1}(-,\Z_n) $ 
is the Bockstein 
 homomorphism associated with the extension
 $\Z_n \stackrel{\cdot m}{\to} \Z_{nm} \to \Z_m$. Thus we get the 5d topological invariant of the $\Spin \times \Z_3$ as:
\begin{equation}
   \exp \big( \ii 2 \pi k  \int_{M^5} 
   ( \frac{1}{18} 
   {\tilde A} (\beta_{(3,3)} {\tilde A}) (\beta_{(3,3)} {\tilde A})
   -\frac{1}{3 \cdot 24} {\tilde A} p_1 ) \big) .
\end{equation}  
 Now when $k = 3$, we get
\begin{equation}
\label{eq:bSPT-Z3}
   \exp \big( \ii 2 \pi   \int_{M^5} 
   ( \frac{1}{6} 
   {\tilde A} (\beta_{(3,3)} {\tilde A}) (\beta_{(3,3)} {\tilde A})
   -\frac{1}{24} {\tilde A} p_1 ) \big)
   =
   \exp \big( \ii 2 \pi   \int_{M^5} 
   ( \frac{1}{6} 
   {\tilde A} (\beta_{(3,3)} {\tilde A}) (\beta_{(3,3)} {\tilde A}))
   \big),
\end{equation}  
while the equality relies on a Theorem of
Tomonaga \cite{tomonaga1964mod}
such that 
$\int_{M^5}  {\tilde A} p_1 = 0 \mod 24$
on the $\Spin \times \Z_3$ manifold.

Next, we show that this $k=3$ class indeed is
the generator of the group cohomology bosonic SPT class in 
$\H^5(\B \Z_3, \U(1))= \Z_3$ tensor product with the free fermion SPT class,
such that the 3 layers of \eq{eq:bSPT-Z3}
becomes a trivial class. Namely, when $k=3 \cdot 3 =9$, we get
\begin{equation}
\label{eq:bSPT-Z3-3}
\exp \big( \ii  \pi   \int_{M^5} 
   {\tilde A} (\beta_{(3,3)} {\tilde A}) (\beta_{(3,3)} {\tilde A})
   \big)
   = \exp \big( \ii  \pi   \int_{M^5} 
   {\tilde A} (4\beta_{(3,3)} {\tilde A}) (\beta_{(3,3)} {\tilde A})
   \big) = 1,
\end{equation} 
which can only read the $0,1 \mod 2$ in 
$\int_{M^5} 
   ({\tilde A} (\beta_{(3,3)} {\tilde A}) (\beta_{(3,3)} {\tilde A}) \mod 2$.
   Now, we use the conflict between the mod 2 index built out of the mod 3 gauge field
   to argue that 
   this is a trivial class of SPTs.
   Note that
$3 (\beta_{(3,3)} {\tilde A})=0 \mod 3$ 
and $(\beta_{(3,3)} {\tilde A})= 4 (\beta_{(3,3)} {\tilde A})= \mod 3$ 
used in the first equality
for the $\Z_3$ valued gauge field, while 
the second equality uses 
$2 (2\pi \Z) = 0 \mod 2 \pi$
thus 
\eq{eq:bSPT-Z3-3} is a trivial class SPTs. 

We may simply consider
$k=3$ case as 
$\exp \big( \ii 2 \pi   \int_{M^5} 
   ( \frac{1}{3} 
   {\tilde A} (\beta_{(3,3)} {\tilde A}) (\beta_{(3,3)} {\tilde A}))
   \big)$
   tensor product with a trivial gapped fermionic state. 
   Then the $k=9$ also generates a $+1$
   as a trivial SPTs,
   This concludes the proof of the group extension in \eq{eq:Z9class}
   as the $\Z_9$ classification of 5d $\Spin \times \Z_3$ topological invariants thus also 5d fermionic SPTs.

\newpage

\bibliography{BSM-TQDM}

\end{document}